# Intrinsic Electron-Phonon Resistivity in $Bi_2Se_3$ in the Topological Regime

Dohun Kim[1], Qiuzi Li[2], Paul Syers[1], Nicholas P. Butch[1†], Johnpierre Paglione[1], S. Das Sarma[1,2], and Michael S. Fuhrer[1*]

1. *Center for Nanophysics and Advanced Materials, Department of Physics, University of Maryland, College Park, MD 20742-4111, USA*

2. *Condensed Matter Theory Center, Department of Physics, University of Maryland, College Park, MD 20742-4111, USA*

**Abstract**

We measure the temperature-dependent carrier density and resistivity of the topological surface state of thin exfoliated $Bi_2Se_3$ in the absence of bulk conduction. When the gate-tuned chemical potential is near or below the Dirac point the carrier density is strongly temperature dependent reflecting thermal activation from the nearby bulk valence band, while above the Dirac point, unipolar *n*-type surface conduction is observed with negligible thermal activation of bulk carriers. In this regime linear resistivity vs. temperature reflects intrinsic electron-acoustic phonon scattering. Quantitative comparison with a theoretical transport calculation including both phonon and disorder effects gives the ratio of deformation potential to Fermi velocity $D/\hbar v_F$ = 4.7 Å$^{-1}$. This strong phonon scattering in the $Bi_2Se_3$ surface state gives intrinsic limits for the conductivity and charge carrier mobility at room temperature of ~550 μS per surface and ~10,000 cm$^2$/Vs.

The 3D topological insulators (TIs) have an insulating bulk but metallic surface states stemming from band inversion due to strong spin-orbit interaction [1, 2]. The helicity of the surface carriers is protected by time reversal symmetry and is exhibited in spin-momentum locking [3] which may be exploited in spintronic [4] and opto-spintronic devices [5], and the discovery of TI materials such as $Bi_2Se_3$ with a bulk bandgap on order 300 meV [2] is anticipated to enable such topological electronic devices at room temperature. While the helical surface state is robust against backscattering by time-reversal-invariant disorder [6], the eventual surface conductivity and mobility is determined entirely by carrier scattering from disorder and phonons since scattering at angles less than $\pi$ is possible in the two-dimensional surface state. The intrinsic limits to conductivity and mobility at finite temperature will be set by the electron-phonon interaction, hence the study of electron-phonon coupling in TIs is of central importance in assessing potential electronic applications. The goal of the current work is to combine experiment and theory in order to ascertain the intrinsic limit to the 2D transport mobility of TI surface carriers arising from phonon scattering.

Experimental studies of inelastic scattering processes at the TI surface have been confined to angle-resolved photoemission [7, 8] or helium atom surface scattering probes [9] of the quasiparticle or phonon lifetimes respectively. These studies do not measure directly the resistivity, and the inferred electron-phonon coupling constants are not consistent with each other, varying by a factor of $\approx$5. The high level of doping in bulk TI crystals, hence dominant bulk conduction, hinders the unambiguous analysis of surface signatures in electronic transport [10, 11]. However we recently demonstrated that for $Bi_2Se_3$ thin ($\approx$10 nm) crystals with low bulk disorder, the top and bottom topological surface states are strongly electrostatically coupled, and their Fermi energies are nearly identical and can be tuned together through the Dirac point by application of only a back-gate voltage, allowing measurement of the carrier-density-dependent electronic transport properties of the topological surface states in the absence of a conducting bulk [12, 13].

Here we report the temperature ($T$) and gate-voltage ($V_g$) dependence of the longitudinal resistivity $\rho_{xx}$ and Hall carrier density $n_H = 1/R_H e$ of the surface states of thin exfoliated $Bi_2Se_3$ crystals in the topological regime, where $R_H$ is the Hall coefficient and $e$ is the elemental charge. $n_H$ is temperature dependent for Fermi energies close to the bulk valence band, and can be well described by thermal activation of carriers from a massive bulk band into the Dirac surface state. For a wide range of carrier densities ($2 \times 10^{12} cm^{-2} < n_H < 8 \times 10^{12} cm^{-2}$; spanning the majority of the range of $n$-type conduction) there is negligible activation of bulk carriers. There we observe $\rho_{xx}$ increasing linearly with temperature, with a slope at high carrier density $\approx 3$ $\Omega$/K. (For the purpose of comparison, this rate is more than an order of magnitude larger than the temperature dependence of the corresponding graphene resistivity [14, 15].) Comparing with theoretical calculations including the effect of disorder-induced charge fluctuations and temperature-dependent screening, we interpret the linear increase in resistivity as reflecting the electron-acoustic phonon scattering in the degenerate metallic surface state. For Fermi energies near the conduction band edge we determine the ratio of deformation potential to Fermi velocity $D/\hbar v_F = 4.7$ Å$^{-1}$. For Fermi velocities of 5-7 x $10^5$ m/s as measured by angle-resolved photoemission (ARPES) at similar Fermi energies [16, 17] the deformation potential is in the range 15-22 eV, in agreement with theoretical expectations.

$Bi_2Se_3$ devices were produced by micro-mechanical cleavage of low-doped (carrier density ~ $10^{17} cm^{-3}$) bulk $Bi_2Se_3$ single crystals with bulk resistivity exceeding 2 m$\Omega$ cm at 300 K [10] on 300nm $SiO_2$/ doped Si substrates, followed by patterning into Hall bar geometry with electron-beam lithography (EBL) and Ar plasma etching. Au/Cr electrodes were defined by EBL (see inset of Fig. 1b). A brief ($\approx$10 s) selective surface treatment of the contact area with $N_2$ or Ar plasma before the deposition of metals was used to enhance Ohmic conduction of the contacts. To avoid band gap opening due to hybridization of bottom and top surfaces [18, 19], crystals with thickness about 10 nm (confirmed by AFM) were used in this study. In order to reduce high *n*-type doping commonly observed in micro-fabricated $Bi_2Se_3$ devices, we employed molecular charge transfer doping by thermal evaporation of 2,3,5,6-tetrafluoro-7,7,8,8-tetracyanoquinodimethane (F4TCNQ) organic molecules on the fabricated devices [12, 20]. After F4TCNQ deposition we typically measure $\underline{n}_H$ in the range 3-5 x $10^{12}$ cm$^{-2}$ at zero gate voltage. The carrier density in the topological surface state at the bulk conduction band edge is approximately 5 x $10^{12}$ cm$^{-2}$ per surface, or 1.0 x $10^{13}$ cm$^{-2}$ total for top and bottom surfaces; hence the devices are in the topological regime before application of a back gate voltage. Temperature-dependent longitudinal and transverse resistivity were measured simultaneously using two lock-in amplifiers and a commercial cryostat equipped with 9 T superconducting magnet.

Figure 1(a) shows $n_H$ as a function of gate voltage $V_g$ and temperature $T$, where several line cuts with respect to $V_g$ at various temperatures are plotted in Fig. 1(b). In general, we observe gapless ambipolar electric field effect which is indicated by the sign change in $n_H$ at the charge neutrality point ($V_{g,0}$, Fig. 1(b) dashed lines) associated with the Dirac point of the topological surface states. The positive (negative) divergence in $n_H$ when approaching $V_{g,0}$ from above (below) is due to inhomogeneity-driven electron-hole puddles [12, 21]. At low temperature (e.g., 2 K), $n_H(V_g)$ is linear over broad ranges of *n*- and *p*-doping with slope d$n_H$/d$V_g$ = $C_g/e$ indicating a gate capacitance $C_g$ = 11 nFcm$^{-2}$ as expected for 300 nm-thick $SiO_2$.

However, as temperature is raised, $V_{g,0}$ shifts to more negative voltage and the slope $dn_H/dV_g$ is reduced near and below $V_{g,0}$, while $n_H(V_g)$ remains temperature-independent for the unipolar $n$-type conduction regime ($V_g > $ -20 V). The shift in $V_{g,0}$ shows thermally activated behavior with activation energy of 60 meV as determined from the Arrhenius fit of $V_{g,0}(T)$ in the range of $50 < T < 150$ K.

We first explore the origin of the temperature dependence of $n_H$. In our previous study [12], we experimentally demonstrated by dual electrostatic gating that significant band bending is absent in thin ($\approx$10 nm), lightly doped $Bi_2Se_3$ crystals due to strong capacitive coupling of the two topological surfaces through the high-dielectric-constant insulating bulk $Bi_2Se_3$. Therefore, although a superficially similar shift of $V_{g,0}$ was observed in Ref. [22] and interpreted there as due to a reduction in the effective bulk gap due to band-bending, we rule out this effect in our devices. Instead, we note that the energy scale for thermal activation corresponds well with the expected separation of the bulk valence band and the Dirac point of the topological surface band. Hence significant thermal activation of electrons from the bulk valence band to the surface band, hence leaving holes in the bulk valence band, should occur at modest temperatures, resulting in a negative shift of $V_{g,0}$.

We develop a more detailed model of $n_H(T)$ as follows. We assume that $n_H$ reflects the density of surface carriers only. Since in a multi-band system the net Hall conductivity is the mobility-weighted average of the Hall conductivities of each band, the assumption reflects the expectation of much lower carrier mobility for the bulk valence band which has much higher effective mass and no topological protection from backscattering. We assume that two bands are important in determining $n_H(T)$: the surface band with dispersion $E_{2D}(k) = \hbar v_{F,0} k + \dfrac{\hbar^2 k^2}{2m_s^*}$ and the bulk valence band with dispersion $E_{3D}(k) = E_v - \dfrac{\hbar^2 k^2}{2m_b^*}$ , where we estimate the energy of band edge $E_v$ = -60 meV from the observed temperature dependent shift of $V_{g,0}$. For the calculation, we used the parameters of the Fermi velocity near Dirac point $v_{F,0}$ = 3 x $10^7$ cm/s, and effective mass $m_s^*$ = 0.3 $m_e$ of the surface band and the effective mass $m_b^*$ = 2.6 $m_e$ of the bulk valence band which are reasonable for $Bi_2Se_3$ from the ARPES measurements [16, 17, 23].

Fig. 1(d) shows $Bi_2Se_3$ density of states $D(E)$ used in our model. Figure 1(c) shows the calculated $n_H(V_g,T)$ based on $D(E)$. The model captures the essential characteristics of the measured $n_H(V_g,T)$ (Fig. 1a): $V_{g,0}$ shifts to negative voltage in a thermally-activated manner, $C_g$ in the hole-doped region is comparable to that of 300nm $SiO_2$ at low temperatures but is gradually suppressed as temperature is increased due to thermal activation from bulk valence band; and $n_H$ in the electron-doped region ($V_g$ > -20 V) is nearly independent of temperature. The model fails to describe the divergence of $n_H$ near $V_{g,0}$ due to charge inhomogeneity which is not included in the model. However, the overall success of the model supports our picture of unipolar $n$-type transport dominated by the topological surface state in the absence of bulk carriers in the region -20 V < $V_{g,0}$ < 80 V, corresponding to carrier densities in the topological surface of 1.5 – 8 x $10^{12}$ cm$^{-2}$.

We now discuss temperature-dependent longitudinal resistivity $\rho_{xx}(T)$ in the unipolar $n$-type regime. Figure 2(a) shows $\rho_{xx}(T)$ of device 1 at different $n_H$ tuned by $V_g$. Generally, $\rho_{xx}(T)$ is metallic ($d\rho_{xx}/dT > 0$) and saturates at low $T$ < 40K as expected for gapless surface states. In the intermediate temperatures of 50 K < $T$ < 150 K, $\rho_{xx}(T)$ is linearly increasing with $T$ consistent with scattering of degenerate electrons with phonons that are populated according to the classical equipartition distribution. For device 1 the highest Hall mobility at low temperature is 1430 cm$^2$/Vs at $n_H = 2.6$ x $10^{12}$cm$^{-2}$ which is reduced by more than 20% to 1120 cm$^2$/Vs at $T = 150$ K. Figure 2(b) shows the average slope d$\rho_{xx}$/d$T$ for 50 K < $T$ < 150 K at different $n_H$ for two different experimental runs with two different samples. d$\rho_{xx}$/d$T$ varies by less than a factor of 1.75 while the carrier density varies almost fourfold. A broad peak in d$\rho_{xx}$/d$T$ occurs around $n_H = 3.5$ x $10^{12}$cm$^{-2}$, and d$\rho_{xx}$/d$T$ approaches a roughly density independent value of $\approx 3$ $\Omega$/K at high $n_H$.

In the high temperature and high carrier density regime, the temperature dependence of resistivity d$\rho_{xx}$/d$T$ mainly comes from the electron-phonon scattering, the asymptotic high-temperature behavior of which can be expressed as

$$\rho_{ph} \sim \frac{\pi D^2 k_B T}{e^2 \hbar v_F^2 v_{ph}^2 \rho_m} \qquad (1)$$

where $\rho_{ph}$ is the electron-phonon scattering limited resistivity, $D$ is the deformation potential and $v_{ph} = 2900$ m/s is the phonon velocity. $\rho_m = 7.68$ x $10^{-7}$ g/cm$^2$ is the mass density of Bi$_2$Se$_3$ for a single quintuple layer (thickness ~1 nm) corresponding to the penetration length of the topological surface state into the bulk [26]. As can be seen from Eq. (1), d$\rho_{ph}$/d$T$ is determined by the ratio of the deformation potential to the Fermi velocity $D/\hbar v_F$.

In real samples with disorder, other effects (e.g. screening, potential fluctuation induced puddles) may contribute to the temperature dependence of the resistivity. Charged impurity scattering and potential fluctuations play important roles in determining the minimum conductivity and low temperature transport of the topological surface state [12, 13, 24]. Due to the strong suppression of the back-scattering in TI surface transport, metallic temperature-dependent screening effects are relatively weak [25], and are therefore not discussed here.

Thermal activation across potential fluctuations gives rise to an insulating behavior of temperature-dependent resistivity ($d\rho/dT < 0$) which may be important at low carrier density where phonon effects are suppressed. To determine whether Eqn. (1) accurately describes the resistivity in our samples, we have developed a realistic model of temperature-dependent resistivity of the topological surface state including electron-phonon scattering [14, 27] and the effects of charged impurity scattering [25] and potential fluctuations [12, 21]. We have calculated the resistivity by using Boltzmann transport theory and effective medium theory, in which we assume the potential fluctuation follows a Gaussian form parameterized by the potential strength *s*. The detailed theories have been presented elsewhere [24, 25].

One serious issue in a quantitative theoretical determination of the temperature dependence is that the deformation potential coupling for the surface electrons is simply not known, and theoretical determination from first principles is imprecise. Since the resistivity depends on the square of the deformation potential $\rho \sim D^2$ any imprecision in *D* is magnified in $\rho$. An experiment-theory comparison is therefore essential to determine the value of *D*. We have therefore carried out a careful regression fit to make the linear *T*-dependence of the experiment and theory agree with each other so as to obtain the best possible estimate of *D* in the system.

Figures 3a and 3b show the calculated $\rho_{xx}(T)$ for two different disorder strengths in the range of temperatures for which the experiment sees linear $\rho_{xx}(T)$. We find the best match to the experimentally measured $d\rho_{xx}/dT$ at high carrier density for a deformation potential $D = 22$ eV and Fermi velocity $v_F = 7 \times 10^5$ m/s ($D/\hbar v_F = 4.7$ Å$^{-1}$). We take the effective background dielectric constant to be $\kappa = 33$, the average distance of the charged impurity from the topological surface $d = 1$ Å. In Fig. 3a, the charged impurity density $n_{imp} = 3 \times 10^{13}$ cm$^{-2}$ (charge inhomogeneity driven potential fluctuation $s = 90$ meV) and in Fig. 3b $n_{imp} = 1.5 \times 10^{13}$ cm$^{-2}$ ($s =$ 65 meV). The disorder strength is the major determinant of the carrier density dependence of the resistivity as reflected in Figs. 3a and 3b. The calculated resistivities bracket the experimental data, indicating that the experimental disorder is comparable to the estimates. A quantitative agreement in the carrier density dependence of the resistivity is not expected since the model does not include the significant non-linearity of the surface bands [13].

Figure 3c shows the calculated slope $d\rho_{xx}/dT$ for 50 K $< T <$ 150 K as a function of carrier density $n$. At low $n$ the slope increases monotonically with density, and is disorder-dependent. At high $n$, the slope asymptotically approachs a constant consistent with Eqn. (1). The results indicate that the slope measured at high $n$ reflects electron-phonon scattering even at large disorder strength; hence the experimentally measured slope of 3 $\Omega$/K determines the ratio $D/\hbar v_F = 4.7$ Å$^{-1}$ at high carrier density. Assuming a Fermi velocity $v_F = 5\text{-}7 \times 10^5$ m/s consistent with ARPES measurements [16, 17, 23], the corresponding deformation potential is $D = 15 - 22$ eV. The experimentally observed $d\rho_{xx}/dT$ shows a peak at lower carrier density, which we understand as a competition between decreasing $v_F$ due to the non-linearity of the surface bands (not included in the theory) which leads to increased $d\rho_{xx}/dT$, and thermal activation across the potential fluctuations which leads to a reduced $d\rho_{xx}/dT$ near the Dirac point as seen in Fig. 3c.

In contrast to the intermediate-$T$ behavior, the resistivity at higher $T >$ 150K is highly non-linear in $T$, and becomes significantly dependent on $n_H$, increasing for decreasing $n_H$ (Fig. 2a). Similar behavior was observed in graphene [15], where the non-linear $\rho_{xx}(T)$ was attributed to remote interfacial scattering of electrons with polar optical phonons in $SiO_2$ substrate. However, fitting high temperature resistivity of $Bi_2Se_3$ devices to the activated form $\rho_{xx}(T) = \rho_0 e^{-\Delta E/k_B T}$ yields 130 meV $< \Delta E <$ 170 meV depending on $n_H$, which appears unreasonable for optical phonons of the substrate or $Bi_2Se_3$. Moreover, we find that the resistivity between 250K < $T$ < 300K shows appreciable hysteresis in temperature, which is likely associated with charge transfer from $SiO_2$ substrates to the bottom surface or absorption and desorption of F4TCNQ molecules on the top surface of the device. Thus we ascribe the observed non-linear behavior at $T > 200K$ to an extrinsic contribution to $\rho_{xx}(T)$ whose origin is not yet understood.

In conclusion, we have studied temperature-dependent electronic transport in the topological surface states of $Bi_2Se_3$. We find significant activation of carriers from the bulk valence band for Fermi energies near or below the Dirac point, but identify a region of unipolar electron conduction where activation of bulk carriers is negligible. In this regime, the temperature-dependence of the resistivity $d\rho_{xx}/dT$ is relatively independent of carrier density, and is used to extract a ratio of deformation potential to Fermi velocity $D/\hbar v_F = 4.7$ Å$^{-1}$ which gauges the strength of the intrinsic electron-acoustic phonon interaction in $Bi_2Se_3$. The strong acoustic phonon scattering places an intrinsic limit on the conductivity of the $Bi_2Se_3$ surface states of ~550 μS = 14 $e^2/h$ per surface, approximately 60 times lower than graphene's intrinsic conductivity at room temperature. The intrinsic mobility limited by electron-acoustic phonon scattering will be carrier-density dependent, and highest at low carrier density. Assuming a factor of two reduction in $v_F$ at the Dirac point (~3 x $10^7$ cm/s) compared to near the bulk conduction band edge (~5-7 x $10^7$ cm/s), the intrinsic carrier density at room temperature is ~2 x $10^{11}$ cm$^{-2}$ and the highest achievable room temperature mobility is <10,000 cm$^2$/Vs. Our combined theoretical-experimental work also provides an accurate estimate for the deformation potential coupling in the $Bi_2Se_3$ surface carriers to be 15 - 22 eV.

\* [mfuhrer@umd.edu](mailto:mfuhrer@umd.edu)

Acknowledgements

The authors acknowledge useful conversations with E. H. Hwang, S. Adam and S. Giraud. This work was supported by NSF grant number DMR-11-05224 and and DARPA QuEST.

Present Address

[†] Condensed Matter and Materials Division, Lawrence Livermore National Laboratory, Livermore, CA 94550, U.S.A.

**Figure captions**

**Figure 1** (a) Measured Hall carrier density ($n_H$) of a thin exfoliated $Bi_2Se_3$ device as a function of back gate voltage ($V_g$) and temperature ($T$). Pink dashed curve shows a $T$ dependent trace of charge neutrality point (CNP; $V_{g,0}$) (b) $n_H$ vs. $V_g$ at various temperatures indicated by dashed lines in (a). Dashed lines indicate points of charge neutrality. The inset shows an optical micrograph of the device. The scale bar is 2μm. (c) Calculated surface band carrier density as a function of $V_g$ and $T$ using a thermal activation model discussed in text. (d) Energy-dependent density of states $D(E)$ for the thermal activation model, showing surface (red and blue solid lines) bands and bulk valence (black solid line) band.

**Figure 2** (a) Longitudinal resistivity $\rho_{xx}$ of device 1 as a function of temperature at different carrier densities tuned by the back gate. (b) Slope of $\rho_{xx}$ vs. carrier density in the linear temperature regime (50 K < $T$ < 150 K). Dashed line indicates the high carrier density value of the slope used to compare to the theoretical model.

**Figure 3** (a,b) Calculated total resistivity $\rho_{xx}$ as a function of temperature for different carrier densities $n$ = 2,3,4,5,6,7,8,9 x $10^{12}$ cm$^{-2}$ from top to bottom. In (a) the charged impurity density is $n_{imp}$ = 3 x $10^{13}$ cm$^{-2}$ and the potential fluctuation strength $s$ = 90 meV. In (b) $n_{imp}$ = 1.5 x $10^{13}$ cm$^{-2}$ and $s$ = 65 meV. (c) Calculated slope d$\rho_{xx}$/d$T$ vs. $n$ in the linear temperature regime (50 K < $T$ < 150 K). The solid line is the result including only electron-phonon scattering, while the dot-dashed and dashed lines correspond to Fig. 3(a) and 3(b), respectively.

Figure 1

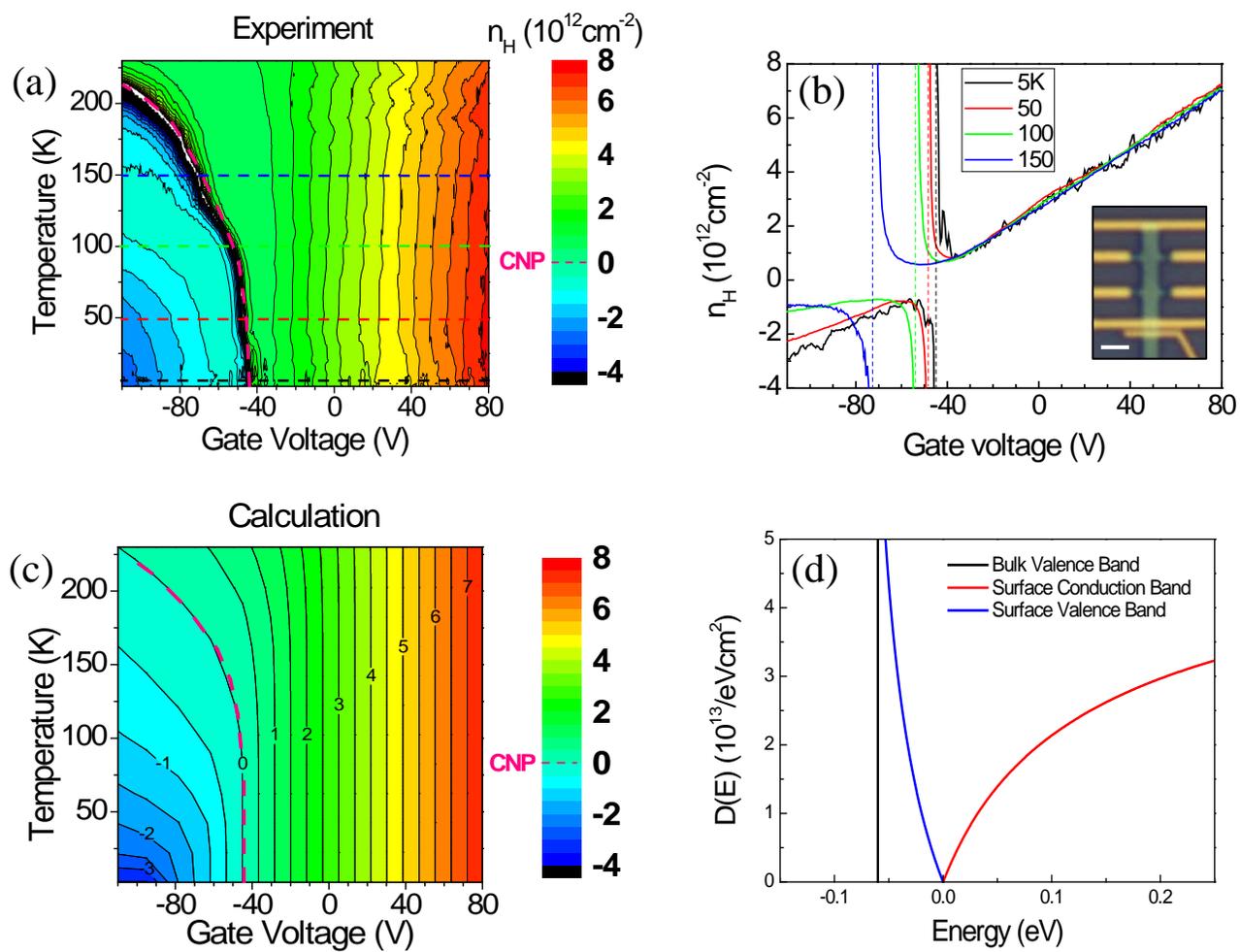

Figure 2

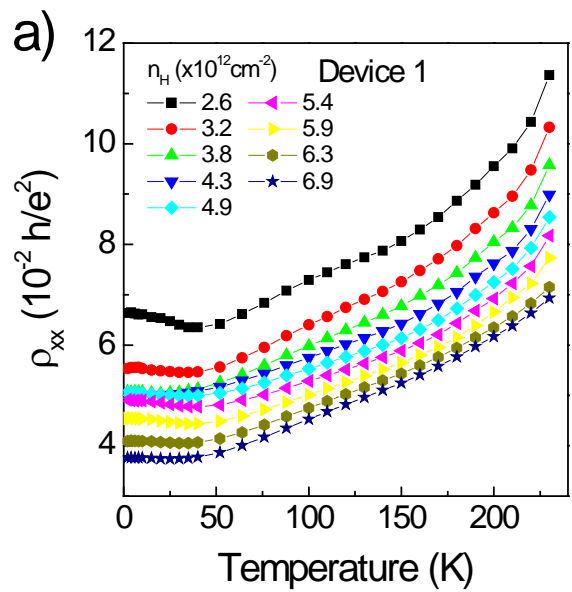

a)

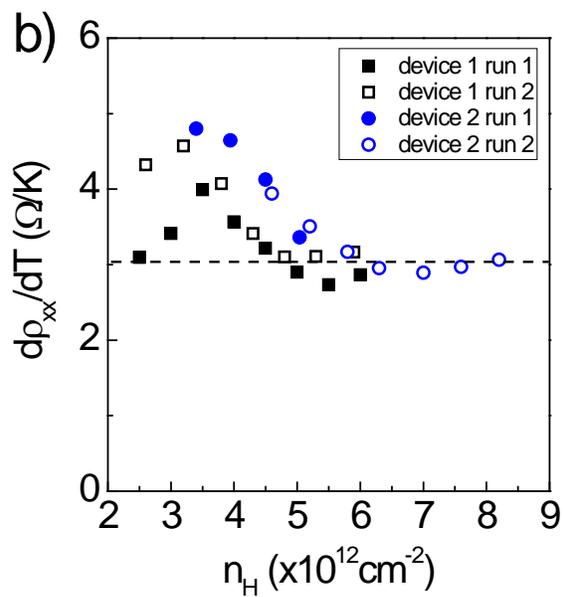

b)

Figure 3

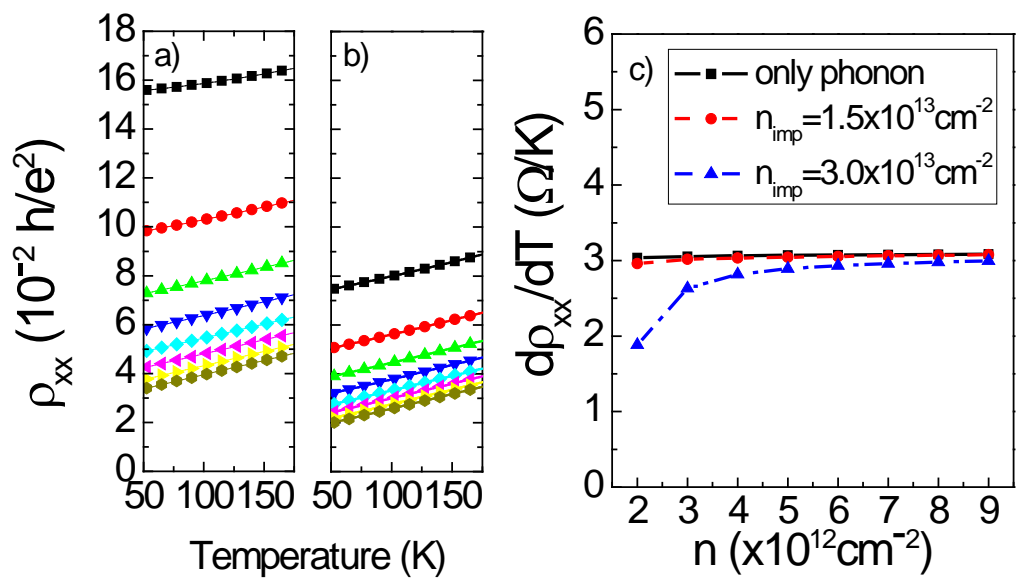